\documentstyle[aps,graphics,epic,eepic]{revtex}
\begin{document}
% \draft command makes pacs numbers print
\draft
\title{Perturbative and Numerical Methods for Stochastic Nonlinear Oscillators}
% repeat the \author\address pair as needed
\author{Giuseppe Curci, Erika D'Ambrosio}
\address{I.N.F.N. Sezione di Pisa, Via Livornese 1291\\ San Piero a Grado I-56010 Pisa Italy}
\address{Dipartimento di Fisica dell'Universit\`a di Pisa \\ P.zza Torricelli 2
, I-56126 Pisa Italy}
\date{29 October 1998}
\maketitle
\begin{abstract}
Interferometric gravitational wave detectors are devoted to pick up the effect
induced on masses by gravitational waves. The variations of the length dividing
two mirrors is measured through a laser interferometric technique. The Brownian
motion of the masses related to the interferometer room temperature is a limit
to the observation of astrophysical signals. It is referred to as thermal noise
and it affects the sensitivity of both the projected and the future generation
interferometers. In this paper we investigate the relevance of small non-linear
effects and point out their impact on the sensitivity curve of interferometric 
gravitational wave detectors (e.g. VIRGO, LIGO, GEO\ldots) through perturbative
methods and numerical simulations.

We find that in the first order approximation the constants characterizing the
power spectrum density (PSD) are renormalized but it retains its typical shape.

This is due to the fact that the involved Feynman diagrams are of tadpole type.

Higher order approximations are required to give rise to up-conversion effects.

This result is predicted by the perturbative approach and is in agreement with
the numerical results obtained by studying the system's non-linear response by
numerically simulating its dynamics.

\end{abstract}
% insert suggested PACS numbers in braces on next line

\section{Introduction}

It is known that every correlation function of a stationary stochastic process 
may be fully obtained as the functional derivative of the associated generator.

This one has to obey some equations connected to the
equations of motion which govern the dynamics of the
system. If we consider a stochastic system and apply
to it all the mathematical techniques and strategies, which have been currently
set up for studying a particle physics problem, then
we get some new ways of approching the Fokker-Planck
equation connected to a random process, both through
perturbative methods and non perturbative ones. This
requires a procedure of quantization starting from a
classical Lagrangian associated to the theory itself
and a passage to the Euclidean metrics.

The probability of shifting from one place to another in the case of a Brownian
motion is expected to obey an evolution equation, like a propagator of Feynman
\footnote{For an exhaustive survey see \cite{SQ,ZJ,FP} and references therein}.

In accordance with that analogy the equation of Fokker-Planck is equivalent to
a Schr\"odinger equation in Euclidean time and the proper Hamiltonian operator
\begin{eqnarray*}
H & = & -D{\partial^2\over{\partial q^2}}+{\partial\over{\partial q}}f(q) \\ &&
-D{\partial^2\over{\partial q^2}}+{1\over2}\left\{\frac{\partial}{\partial q},
f(q)\right\}+{1\over2}\left[\frac{\partial}{\partial q},f(q)\right] \\ &&
-D{\partial^2\over{\partial q^2}}+{1\over2}\left\{\frac{\partial}{\partial q},
f(q)\right\}+{1\over2}f'(q)\qquad.
\end{eqnarray*}

This means that it is possible to study a stochastic process using the methods
of quantum mechanics. Let's introduce the \`a la Schwinger functional \cite{JS}
\begin{eqnarray*}
Z(J,K) &=& \int<q,t_f|{\cal U}_{JK}(t_f,t_i)|q_i,t_i>\mbox{d}q \\ {\cal U}_{JK}
(t_f,t_i) &=&\mbox{T}\exp\int^{t''}_{t'}(J(t)\hat q(t)-iK(t)\hat p(t))\mbox{d}t
\end{eqnarray*}
with $\hat q(t)$ and $\hat p(t)$ taken as operators in the Heisenberg's picture.
The equations of motion
\begin{eqnarray*}
{\mbox{d}\over{\mbox{d}t}}{\delta Z(J,K)\over{\delta J(t)}}&=& [K(t)+2D{\delta
\over{\delta K(t)}}+f({\delta\over{\delta J(t)}})]Z(J,K)\\ {\mbox{d}\over{\mbox
{d}t}}{\delta Z(J,K)\over{\delta K(t)}}&=& [-J(t)-f'({\delta\over{\delta J(t)}}
){\delta\over{\delta K(t)}}+{1\over 2}f''({\delta\over{\delta J(t)}})]Z(J,K).
\end{eqnarray*}
are known as Schwinger-Dyson equations \cite{SQ} and rarely they are completely
solvable. It has been shown in the literature that approximate solutions may be
yielded by expanding order by order the above equations if the leading term in
the drift force is linear. Indeed $Z(J,K)$ has the form of a path-integral with
\begin{equation}
L=\sum_{a,b}(4D)^{-1}_{ab}(\dot q_a-f_a)(\dot q_b-f_b)+
\sum_a{1\over2}\partial_af_a 
\end{equation}
the Lagrangian function in the general case of more than one dimension. What is
the physical r\^ole of $D$? It stands for the noise's amplitude and if it tends
to zero we recover the classical macroscopic equations without fluctuations. It
is the same situation as in quantum mechanics where the uncertainty is related
to the constant $\hbar$. Exploiting this analogy we find an equivalence between
\begin{eqnarray*}
\left[{\mbox{d}\over{\mbox{d}t}}{\delta\over{\delta
J_a(t)}}-f_a({\delta\over{\delta J(t)}})\right]Z(J,
K) & = & [2D_{ab}{\delta\over{\delta K_b(t)}}+K_a(t)
]Z(J,K) \\ \left[{\mbox{d}\over{\mbox{d}t}}
\delta_{ab}+\partial_a f_b({\delta\over{\delta
J(t)}})\right]{\delta Z(J,K)\over{\delta K_b(t)}}
& = & [-J_a(t)+{1\over 2}\partial_a\partial_b
f_b({\delta\over{\delta J(t)}})]Z(J,K) \quad.
\end{eqnarray*}
and the Heisenberg equations.

Starting from the lowest order approximation one may solve the Schwinger-Dyson
equations above with an iterative method. The next step is using the functional
$$Z(J,K)=\exp W(J,K)$$
to recover from $W(J,K)$ the connected correlation functions. On the other hand
it may be expected that the same averages may be obtained following a Langevin
approach that is studying a stochastic differential equation whose solution is
distributed with $P(q,t|q_0,t_0)$ satisfying the Fokker-Planck equation. It may
be assumed that after a transient period the system becomes stationary and its
statistical properties are independent from the initial values $q_0$ and $t_0$.

\section{Mathematical Formalism}

A physical system is usually represented by many deegres of freedom with their
characteristic evolution times. Those variables that evolve very rapidly may be
integrated over and their contribution is averaged over long time periods. What
we are interested in is the resulting evolution of the system. The influence of
the microscopic deegres of freedom on the macroscopic ones may be expressed as 
a term of interaction \cite{Sa,Ku}.

The characteristic times of such interaction are very short in comparison with
the characteristic times of the slowly evolving deegres of freedom and the two 
time scales may be treated separately, at the expense of including a stochastic
term in the equation of motion written for the macroscopic variables. Such term
must be interpreted as follows. At any time there are so many interactions that
we may only treat them by their statistical properties and consider the effect
on the macroscopic system as little fluctuations over the deterministic motion.

Another effect is that phenomenological constants appearing in the equation of
motion are related to the interactions with the microscopic deegres of freedom.

So it is only a matter of convenience separating deterministic from stochastic
contributions in the equation \cite{Fo,VK}.

As an example we treat a classical harmonic oscillator subjected to stochastic
forces. It is convenient to use the notation of the Langevin equation that is a
common equation of motion with a random contribution. The friction constant and
the elastic constant are $2m\xi\omega_0$ and $m\omega_{0}^{2}$, $m$ is the mass
and $D$ is the diffusion constant whose relation with $K_BT$ we do not use yet.

The evolution equation is
$$m\ddot q(t)+2m\xi\omega_0\dot q(t)+m\omega_{0}^{2}q(t)=F(t)\qquad<F(t)F(t')>=
2D\delta(t-t')\, .$$
The familiar method for solving equations of that type is by means of a vector
$$\dot\Phi=-\Gamma\Phi+F\qquad\qquad\Phi=(\begin{array}{c}q \\ p \end{array})$$
with the following notation
$$p=m\dot q\qquad\omega_R=\omega_0\sqrt{1-\xi^2}\qquad,\qquad
\Gamma=(\begin{array}{cc}0&-{1\over m} \\ m\omega_{0}^{2} & 2\xi\omega_0
\end{array})\quad.$$
There are two linearly independent eigenvectors characterized by the condition
$$p=-m\omega_0(\xi\pm i{\omega_R\over\omega_0})q\quad.$$
By the similarity transformation
$$\Gamma'=M^{-1}\Gamma M\quad\quad M={1\over\sqrt{1-(\xi+i{\omega_R\over
\omega_0})^2}}(\begin{array}{cc} 1 & {-1\over{m\omega_0}}(\xi+i{\omega_R\over
\omega_0})\\ -m\omega_0(\xi+i{\omega_R\over\omega_0}) & 1 \end{array})$$
the state vector may be expressed in terms of the eigenvectors of $\Gamma$. The
eigenvalues we get are
$$\Gamma'=(\begin{array}{cc}\omega_0(\xi+i{\omega_R\over\omega_0}) & 0 \\ 0 &
\omega_0(\xi-i{\omega_R\over\omega_0})\end{array})=(\begin{array}{cc}\lambda &
0 \\ 0 &\lambda* 
\end{array})\quad.$$
Formally the solution of the Langevin equation for given initial conditions is
\begin{equation}
\Phi'(t)=e^{-\Gamma't}\Phi'(0)+\int_{0}^{t}e^{\Gamma'(\tau-t)}F'(\tau)d\tau
\label{eq:ph}
\end{equation}
where we have introduced
$$\Phi'=M^{-1}\Phi={1\over\sqrt{1-(\xi+i{\omega_R\over\omega_0})^2}}
\left(\begin{array}{c}q+{1\over{m\omega_0}}(\xi+{i\omega_R\over\omega_0})p \\ 
m\omega_0(\xi+i{\omega_R\over\omega_0})q+p\end{array}\right)$$
$$
F'=M^{-1}(\begin{array}{c} 0 \\ F(t)\end{array})={1\over\sqrt{1-(\xi+i{\omega_R
\over\omega_0})^2}}\left(\begin{array}{c}{1\over{m\omega_0}}(\xi+i{\omega_R
\over\omega_0})F(t) \\ F(t)
\end{array}\right)
$$
as state vector and forcing term in the basis of eigenvectors of $\Gamma$. This
approach seems rather artificial but it is justified by the problem of writing
the evolution operator. As a consequence of the transformation we applied above
the evolution operator becomes $e^{-\Gamma't}$ and its form is now very simple.

The solutions of the Langevin equation are merely linear superpositions of the
eigenvectors of $\Gamma$. The correlation functions for the stochastic term are
\begin{equation}
<F'^{*}_{i}(t)F'_j(t')>=2D'_{ij}\delta(t-t')\label{eq:ft}
\end{equation}
and it should be pointed out that the diffusion constant changes into a tensor
\cite{Th}.

The explicit form is
$$D'_{ij}={D\omega_0\over{2\omega_R}}(\begin{array}{cc}{1\over(m\omega_0)^2} & 
{1\over{m\omega_0}}(\xi-i{\omega_R\over\omega_0}) \\ {1\over{m\omega_0}}(\xi+
{i\omega_R\over\omega_0}) & 1 \end{array})$$
and as a consequence
$$<\Phi'^{*}_{i}(t)\Phi'_j(t')>={De^{-\xi\omega_0|t-t'|}\over{2\xi\omega_R(m
\omega_0)^2}}(\begin{array}{cc}e^{i\omega_R(t-t')} & m\xi\omega_0e^{i
\omega_R|t-t'|} \\ m\xi\omega_0e^{-i\omega_R|t-t'|} &
(m\omega_0)^2e^{-i\omega_R(t-t')} \end{array})\quad.$$
Since we are interested in stationary properties $t$ ad $t'$ are very large. If
this were not true the initial conditions should be taken into account. Even if
we preserve the terms containing $\Phi'(0)$ they all vanish for large times. An
intuitive idea may be drawn by considering the two components in (\ref{eq:ph}).

The term in $\Phi'(0)$ doesn't correlate with anything. When we take an average
using the properties (\ref{eq:ft}) the terms in $\Phi'(0)$ are simply a factor.

When times increase they tend faster to zero than possible averaged terms they 
are multiplied for. The correlation functions written above may be turned again
in the original basis
\begin{eqnarray*}
<q(t)q(t')>& = &{De^{-\xi\omega_0|t-t'|}\over{2\xi m^2\omega_{0}^{3}}}\left(
\cos\omega_R(t-t')+{\xi\omega_0\over\omega_R}\sin\omega_R|t-t'|\right) \\
<q(t)p(t')>&=&{De^{-\xi\omega_0|t-t'|}\over{2m\xi\omega_0\omega_R}}\sin\omega_R
(t-t') \\ <p(t)q(t')>&=&{-De^{-\xi\omega_0|t-t'|}\over{2m\xi\omega_0\omega_R}}
\sin\omega_R(t-t') \\
<p(t)p(t')>&=&{D\over{2\xi\omega_0}}e^{-\xi\omega_0|t-t'|}\left(\cos\omega_R
(t-t')-{\xi\omega_0\over\omega_R}\sin\omega_R|t-t'|\right)\quad.
\end{eqnarray*}
When we have a connection of several oscillators the procedure stated above is
the simplest one because of the dimensionality of $\Gamma$. For now it may seem
a tortuous mathematical way.

\subsection{Generating Functionals}

It may be shown that the results of the preceding subsection coincide with the
treatment in the introduction.

This is an example in which an analytic solution is available. If $\Gamma$ is
a matrix whose eigenvalues are distinct and non zero we may use the convention
\begin{eqnarray*}
\Gamma_{ij}\Phi'^{(\alpha)}_{j}=\lambda_\alpha\Phi'^{(\alpha)}_{i} &\qquad& 
\tilde\Phi'^{(\alpha)}_{i}\Gamma_{ij}=\lambda_\alpha\tilde\Phi'^{(\alpha)}_{j}
\\ \sum_{\alpha}\Phi'^{(\alpha)}_{i}\tilde\Phi'^{(\alpha)}_{j}=\delta_{ij}
&\qquad&
\sum_i\Phi'^{(\alpha)}_{i}\tilde\Phi'^{(\beta)}_{i}=\delta_{\alpha\beta}
\end{eqnarray*}
with the Hamiltonian operator
\begin{equation}
\hat H =\sum_{i,j}D_{ij}\hat{\Pi}_i\hat{\Pi}_j+i\sum_j\hat{\Pi}_j f_j(\hat\Phi)
\qquad f_i(\Phi)=-\Gamma_{ij}\Phi_j \qquad.
\end{equation}
It may be shown that the related Schwinger-Dyson equations we should solve are
\begin{eqnarray*}
({\mbox{d}\over{\mbox{d}t}}\delta_{ij}+\Gamma_{ij}){\delta W(J,K)\over{\delta
J_j(t)}} & = & K_i(t)+2D_{ij}{\delta W(J,K)\over{\delta K_j(t)}} \\ ({\mbox{d}
\over{\mbox{d}t}}\delta_{ij}-\Gamma_{ji}){\delta W(J,K)\over{\delta K_j(t)}}&=&-J_i(t)
\end{eqnarray*}
from which we derive
\begin{eqnarray*}
({\mbox{d}\over{\mbox{d}t}}\delta_{ij}+\Gamma_{ij})
{\delta^2W(J,K)\over{\delta J_j(t)\delta K_h(\tau)}}
& = & \delta_{ih}\delta(t-\tau) \\
({\mbox{d}\over{\mbox{d}t}}\delta_{ij}-\Gamma_{ji})
{\delta^2W(J,K)\over{\delta K_j(t)\delta J_h(\tau)}}
& = & -\delta_{ih}\delta(t-\tau) \\
({\mbox{d}\over{\mbox{d}t}}\delta_{ij}+\Gamma_{ij})
{\delta^2W(J,K)\over{\delta J_j(t)\delta J_h(\tau)}}
& = & 2D_{ij}{\delta^2W(J,K)\over{\delta K_j(t)
\delta J_h(\tau)}} \end{eqnarray*}
and the solutions are
\begin{eqnarray*}
{\delta^2W(J,K)\over{\delta J_i(t)\delta K_j(\tau)}}&=&\theta(t-\tau)[e^{-\Gamma(t-\tau)}]_{ij}=\theta(t-\tau)\sum_\alpha e^{-\lambda_\alpha(t-\tau)}
\Phi'^{(\alpha)}_{i}\tilde\Phi'^{(\alpha)}_{j} \\
\frac{\delta^2W(J,K)}{\delta J_i(t)\delta J_j(\tau)}&=&\sum_{\alpha,\beta}
\frac{2}{\lambda_\alpha+\lambda_\beta}\Phi'^{\alpha}_{i}\tilde D^{\alpha\beta}
\Phi'^{\beta}_{j}[e^{\frac{\lambda_\alpha-\lambda_\beta}{2}(\tau-t)-
\frac{\lambda_\alpha+\lambda_\beta}{2}|\tau-t|}-e^{-\lambda_\alpha(t-t_0)}
e^{-\lambda_\beta(\tau-t_0)}] 
\end{eqnarray*}
where we have introduced
\begin{equation}
\tilde D^{\alpha\beta}=
\tilde\Phi'^{(\alpha)}_{i}D_{ij}\tilde\Phi'^{(\beta)}_{j}
\end{equation}
and $t_0$ and $t_f$ are the initial and final times. Using the above results we
may write the functional
$$W(J,K)=\int_{t_0}^{t_f}\int_{t_0}^{t_f}J_i(t)
{\delta^2W(J,K)\over{\delta J_i(t)\delta K_j(\tau)}}
K_j(\tau)\mbox{d}t\mbox{d}\tau+\int_{t_0}^{t_f}
\int_{t_0}^{t_f}J_i(t){\delta^2W(J,K)\over{\delta
J_i(t)\delta J_j(\tau)}}J_j(\tau)\mbox{d}t\mbox{d}\tau$$
if the reference system is chosen in such a way that $\Phi_i(t_0)=0$. Otherwise
it appears a term linear in $J$ that takes the initial conditions into account.

Hence we have constructed a solution for the connected correlations' generator.

\subsection{Non-linear Corrections}

Now suppose that the response of the oscillator is not perfectly linear. We may
write the differential equation
$$m\ddot q(t)+2m\xi\omega_0\dot q(t)+m\omega_{0}^{2}q(t)+\epsilon q^3(t)=F(t)$$
to describe such a situation. The parameter $\epsilon$ is assumed so small that
the system's behaviour can be described by a perturbative approach. On a formal
level we are searching for the eigenvectors of the evolution operator that have
the property of being non-interacting eigenstates of the system. Because of the
perturbative term in the equation of motion, some new cross-terms appear in the
evolution operator whose representative matrix is no more diagonal, but for the
description of the system the basis of imperturbed eigenvectors is still valid
\cite{Na,GH}.

The estimated corrections are
\begin{eqnarray*}
\Delta<\Phi'^{*}_{1}(t)\Phi'_1(t')> & = & {-3D^2\epsilon e^{-\Re\lambda|t-t'|}
e^{i\Im\lambda(t-t')}\over{2m^3(2m\xi\omega_0)^2|\lambda|^5\Im\lambda}}\left[1+
|\lambda|^2{t-t'\over{i\Im\lambda}}+\Re\lambda{1-e^{-2i\Im\lambda(t-t')}
\over{2\Im\lambda}}({\Re\lambda\over{\Im\lambda}}-i{t-t'\over{|t-t'|}})
\right] \\
\Delta<\Phi'^{*}_{1}(t)\Phi'_2(t')> & = & {-3D^2\epsilon\over{2m^2(2m\xi
\omega_0)^2
|\lambda|^3(\Im\lambda)^2}}e^{-\lambda^*(|t-t'|)}\left[{\lambda-\lambda^*e^{-2i
\Im\lambda|t-t'|}\over{2\Im\lambda}}-i\Re\lambda({1\over\lambda^*}+|t-t'|)
\right] \\
\Delta<\Phi'^{*}_{2}(t)\Phi'_1(t')> & = & {-3D^2\epsilon\over{2m^2(2m\xi\omega_0)^2
|\lambda|^3(\Im\lambda)^2}}e^{-\lambda|t-t'|}\left[{\lambda^*-\lambda e^{2i\Im
\lambda|t-t'|}\over{2\Im\lambda}}+i\Re\lambda({1\over\lambda}+|t-t'|)\right] \\
\Delta<\Phi'^{*}_{2}(t)\Phi'_2(t')> & = & {-3D^2\epsilon e^{-\Re\lambda|t-t'|}
e^{-i\Im\lambda(t-t')}\over{2m(2m\xi\omega_0)^2|\lambda|^3\Im\lambda}}[1-
|\lambda|^2{t-t'\over{i\Im\lambda}}+ Re\lambda{1-e^{2i\Im\lambda(t-t')}\over{2
\Im\lambda}}({\Re\lambda\over{\Im\lambda}}+i{t-t'\over{|t-t'|}})]
\quad.
\end{eqnarray*}
In the physical basis
\begin{eqnarray*}
\Delta<q(t)q(t')> & = & {3\epsilon D^2e^{-\Re\lambda|t-t'|}\over{8m^5
|\lambda|^4(\Re\lambda)^2}}[({|t-t'|\over{(\Im\lambda)^2\over{\Re\lambda}}
}-{2\over|\lambda|^2})\cos\Im\lambda(t-t')-[{|t-t'|\over{\Im\lambda}}+{3\!-2\!(
{\Re\lambda\over|\lambda|}]^2\over{\Im\lambda|\lambda|^2\over{\Re\lambda}}})
\sin\Im\lambda|t-t'|] \\
\Delta<q(t)p(t')>& = &{-3\epsilon D^2 e^{-\Re\lambda|t-t'|}\over{8m^4
|\lambda|^2(\Re\lambda)^2(\Im\lambda)^2}}\left[{1\over\Im\lambda}\sin\Im\lambda
(t-t')-(t-t')\cos\Im\lambda(t-t')\right] \\
\Delta<p(t)q(t')> &=& {3\epsilon D^2 e^{-\Re\lambda|t-t'|}\over{8m^4|\lambda|^2
(\Re\lambda)^2(\Im\lambda)^2}}\left[{1\over\Im\lambda}\sin\Im\lambda(t-t')-(t-
t')\cos\Im\lambda(t-t')\right] \\
\Delta<p(t)p(t')> &=& {-3\epsilon D^2e^{-\Re\lambda|t-t'|}\over{8m^3|\lambda|^2
\Re\lambda(\Im\lambda)^2}}\left[|t-t'|\cos\Im\lambda(t-t')+({\Im\lambda\over{
\Re\lambda}}|t-t'|-{1\over{\Im\lambda}})\sin\Im\lambda|t-t'|\right]
\end{eqnarray*}
are the correlation functions.

These relations require no knowledge of $t$ and $t'$ but only of $t-t'$. If $t$
and $t'$ were not very large this property should no longer be true. Throughout
we use infinitely large values for times, to calculate correlation functions in
equilibrium conditions. In the stationary limit we may write the power spectrum
density (PSD) as follows
\begin{equation}
S(f)=\int_{-\infty}^{+\infty}e^{-2\pi if\tau}<q(t)q(t+\tau)>\mbox{d}\tau
\label{eq:sf}
\end{equation}
since $<q(t)q(t+\tau)>$ is time-independent and depends on $\tau$ only. We have
\begin{equation}
S(f)={D\over{2(m\xi\omega_0)^2}}\frac{1}{({\omega_0\over{2\xi}}-{2\pi^2f^2\over{\xi\omega_0}})^2+4\pi^2f^2}-\frac{12\epsilon{D^2\over{(m\omega_{0}^{2})^2
(2m\xi\omega_0)^4}}({\omega_0\over{2\xi}}-{2f^2\pi^2\over{\xi\omega_0}})}{[({
\omega_0\over{2\xi}}-{2\pi^2f^2\over{\xi\omega_0}})^2+4\pi^2f^2]^2}
\end{equation}
which can be written
\begin{equation}
S(f)={2D\over m^2}\frac{1}{(\omega_{0}^{2}(1+{3D\epsilon\over{2\xi m^3
\omega_{0}^5}}-4\pi^2f^2)^2+(4\pi f m\xi\omega_0)^2}
\end{equation}
in the first order approximation. Now we have several ways of interpreting this
result. We have searched for stationary solutions of the Langevin equation when
non-linear terms appear. These terms are related to non-quadratic contributions
in the Lagrangian function \cite{ED}.

This means that in the equilibrium state the proper frequency changes with the
local form of the potential function. Consequently the parameters governing the
responsive behaviour of the system appear modified. As outlined above the term 
in $\epsilon$ must contribute very little in the motion equation. Without going
into details we say that $m\omega_{0}^{2}q>>\epsilon q^3$. Since in equilibrium
conditions $<q^2>\sim{D\over{2\xi m^2\omega_{0}^{3}}}$ the estimated provision
${D\epsilon\over{2\xi m^3\omega_{0}^{5}}}<<1$ must be satisfied. This parameter
is the same appearing in the corrections recovered in this section. They should
be wrong if ${D\epsilon\over{2\xi m^3\omega_{0}^{5}}}<<1$ were not provided. If
the system is in its transient state $<q^2>$ depends on its initial conditions
and its value may be larger than the constant ${m\omega_{0}^{2}\over\epsilon}$.

In such a situation we cannot use perturbative methods to study the non-linear
component of the force.

\subsection{Renormalized Coefficients}

Suppose that the oscillator is in contact with a bath and that the probability
distribution in the asymptotic limit is related to the Boltzmann statistics. We
shall show that the dynamic correlation functions we obtained and the averages
we are getting in the next calculations are consistent. In the stationary limit
\begin{eqnarray*}
<q^2>=\frac{\int_{-\infty}^{\infty}\int_{-\infty}^{\infty}e^{-{1\over{K_BT}}
({p^2\over{2m}}+{1\over2}m\omega_{0}^{2}q^2+\epsilon{q^4\over4})}q^2\mbox{d}q
\mbox{d}p}{\int_{-\infty}^{\infty}\int_{-\infty}^{\infty}e^{-{1\over{K_BT}}
({p^2\over{2m}}+{1\over2}m\omega_{0}^{2}q^2+\epsilon{q^4\over4})}\mbox{d}q
\mbox{d}p}={K_BT\over{m\omega_{0}^{2}}}(1-{3\epsilon K_BT\over(m\omega_{0}^{2})^2}) &\qquad& t\rightarrow\infty
\\ <p^2>=\frac{\int_{-\infty}^{\infty}\int_{-\infty}^{\infty}e^{-{1\over{K_B
T}}({p^2\over{2m}}+{1\over2}m\omega_{0}^{2}q^2+\epsilon{q^4\over4})}p^2
\mbox{d}q\mbox{d}p}{\int_{-\infty}^{\infty}\int_{-\infty}^{\infty}e^{-{1\over
{K_BT}}({p^2\over{2m}}+{1\over2}m\omega_{0}^{2}q^2+\epsilon{q^4\over4})}
\mbox{d}q\mbox{d}p}=mK_BT
&\qquad&t\rightarrow\infty
\end{eqnarray*}
in the first order approximation. If we call $D=2m\xi\omega_{0}K_BT$ we obtain
the same values estimated above for \\
$<q(t)q(t+\tau)>$ and $<p(t)p(t+\tau)>$ if $\tau=0$.

We may conclude that
$$<q^2>=\frac{\int_{-\infty}^{\infty}\int_{-\infty}^{\infty}e^{-{H\over{K_BT
}}}q^2\mbox{d}q\mbox{d}p}{\int_{-\infty}^{\infty}\int_{-\infty}^{\infty}e^{-{H
\over{K_BT}}}\mbox{d}q\mbox{d}p}\quad\quad<p^2>=\frac{\int_{-\infty}^{\infty
}\int_{-\infty}^{\infty}e^{-{H\over{K_BT}}}p^2\mbox{d}q\mbox{d}p}{\int_{-\infty
}^{\infty}\int_{-\infty}^{\infty}e^{-{H\over{K_BT}}}\mbox{d}q\mbox{d}p}$$
are the same with
$$H={p^2\over{2m}}+{1\over2}m\omega_{0}^{2}q^2+\epsilon{q^4\over4}\qquad
\mbox{or}\qquad
H={p^2\over{2m}}+{1\over2}m\omega_{0}^{2}(1+3\epsilon{K_BT\over{m^2
\omega_{0}^{4}}})q^2$$
in the first order approximation. Finally the relation $D=2m\xi\omega_0K_BT$ is
a thermodynamic property which we expect by the fluctuation-dissipation theorem
\cite{CW}.

Wherever there is damping there must be fluctuations. They are small because of
the factor $K_BT$. Such a size is known beforehand because the fluctuations are
related to ordinary equilibrium statistical mechanics. From a macroscopic point
of view the fluctuation-dissipation theorem tells us that the equilibrium is a
sort of balance between two opposing tendencies (the noise and the damping) and
that the fluctuations are determined by the temperature. This fact is expressed
by the identity $D=2m\xi\omega_0K_BT$ for the macroscopic parameters. $F(t)$ is
treated as an external force whose stochastic properties are given by both the
constants $2m\xi\omega_0$ and $T$ and it acts regardless of $q$. Even if $F(t)$
is irregular and unpredictable we may fully express its properties in terms of
the postulated autocorrelation function.
Then the differential equation
$$m\ddot q(t)+2m\xi\omega_{0}\dot q(t)+m\omega_{0}^{2}q(t)=F(t)\quad\quad
<F(t)F(t')>=2D\delta(t-t')$$
is well-defined with a well-defined solution. Its characteristic Green function
is obtained from inserting
\begin{equation}
F(t)=\delta(t)\qquad\mbox{and has the form}\qquad{\cal G}(t)={\theta(t)\over m}
e^{-\xi\omega_0t}\frac{\sin\omega_Rt}{\omega_R}
\end{equation}
where we intended $\omega_R=\omega_0\sqrt{1-\xi^2}$ as before. Let us introduce
$$m\ddot q(t)+2m\xi\omega_0\dot q(t)+m\omega_{0}^{2}q(t)+\epsilon q^3(t)=F(t)
\qquad.$$
This Langevin equation defines
\begin{eqnarray*}
q(t) & = &\int_{-\infty}^{+\infty}{\cal G}(t-\tau)F(\tau)\mbox{d}\tau-\epsilon
\int_{-\infty}^{+\infty}{\cal G}(t-\tau)q^3(\tau)\mbox{d}\tau \\ & = &
\int_{-\infty}^{+\infty}{\cal G}(t-\tau)F(\tau)\mbox{d}\tau-\epsilon
\int_{-\infty}^{+\infty}{\cal G}(t-\tau)\left[\int_{-\infty}^{+\infty}{\cal G}
(\tau-\tau')F(\tau')\mbox{d}\tau'-\epsilon\int_{-\infty}^{+\infty}
{\cal G}(\tau-\tau')q^3(\tau')\mbox{d}\tau'\right]^3\mbox{d}\tau \\ & = & 
\int_{-\infty}^{+\infty}{\cal G}(t-\tau)F(\tau)\mbox{d}\tau-\epsilon
\int_{-\infty}^{+\infty}{\cal G}(t-\tau)\left[\int_{-\infty}^{+\infty}{\cal G}
(\tau-\tau')F(\tau')\mbox{d}\tau'\right. \\ & & \left. -\epsilon
\int_{-\infty}^{+\infty}{\cal G}(\tau-\tau')[\int_{-\infty}^{+\infty}{\cal G}
(\tau'-\tau'')F(\tau'')\mbox{d}\tau''-\epsilon\int_{-\infty}^{+\infty}
{\cal G}(\tau'-\tau'')q^3(\tau'')\mbox{d}\tau'']^3\mbox{d}\tau'\right]\mbox{d}
\tau \\ & = & \ldots
\end{eqnarray*}
which is clearly a series in $\epsilon$ as it should. For fixed small values of
$\epsilon$ we may consider the first terms only. The initial point is $q(t)$ if
$\epsilon=0$. It has an autocorrelation function $<q(t)q(t')>=R(|t-t'|)$. It is
permissible to use the expansion in $\epsilon$ in order to get the corrections
\begin{eqnarray*}
\Delta<q(t)q(t')>&=&-3\epsilon\int_{-\infty}^{+\infty}{\cal G}(t-\tau)
R(|t'-\tau|)R(0)\mbox{d}\tau-3\epsilon\int_{-\infty}^{+\infty}{\cal G}
(t'-\tau)R(|t-\tau|)R(0)\mbox{d}\tau \\ & &
+18\epsilon^2\int_{-\infty}^{+\infty}{\cal G}(t-\tau)R(|t'-\tau|)
\int_{-\infty}^{+\infty}{\cal G}(\tau-\tau')R(|\tau-\tau'|)R(|\tau-\tau'|)R(0)
\mbox{d}\tau\mbox{d}\tau' \\ & &
+18\epsilon^2\int_{-\infty}^{+\infty}{\cal G}(t'-\tau)R(|t-\tau|)
\int_{-\infty}^{+\infty}{\cal G}(\tau-\tau')R(|\tau-\tau'|)R(|\tau-\tau'|)R(0)
\mbox{d}\tau\mbox{d}\tau' \\ & &
+9\epsilon^2\int_{-\infty}^{+\infty}{\cal G}(t-\tau)R(0)\int_{-
\infty}^{+\infty}{\cal G}(\tau-\tau')R(0)R(|t'-\tau'|)\mbox{d}\tau'\mbox{d}\tau
\\ & &
+9\epsilon^2\int_{-\infty}^{+\infty}{\cal G}(t'-\tau)R(0)\int_{-
\infty}^{+\infty}{\cal G}(\tau-\tau')R(0)R(|t-\tau'|)\mbox{d}\tau'\mbox{d}\tau
\\ & & +18\epsilon^2\int_{-\infty}^{+\infty}{\cal G}(t-\tau)
\int_{-\infty}^{+\infty}{\cal G}(\tau-\tau')R(|\tau-\tau'|)^2R(|t'-\tau'|)
\mbox{d}\tau'\mbox{d}\tau \\ & &  +18\epsilon^2
\int_{-\infty}^{+\infty}{\cal G}(t'-\tau)\int_{-\infty}^{+\infty}{\cal G}(\tau-
\tau')R(|\tau-\tau'|)^2R(|t-\tau'|)\mbox{d}\tau'\mbox{d}\tau\\ & & +9\epsilon^2
\int_{-\infty}^{+\infty}{\cal G}(t-\tau)\int_{-\infty}^{+\infty}{
\cal G}(t'-\tau')R(0)^2R(|\tau-\tau'|)\mbox{d}\tau'\mbox{d}\tau\\ & &
+6\epsilon^2\int_{-\infty}^{+\infty}{\cal G}(t-\tau)
\int_{-\infty}^{+\infty}{\cal G}(t'-\tau')R(|\tau-\tau'|)^3\mbox{d}\tau'
\mbox{d}\tau \qquad.
\end{eqnarray*}

\section{Non-linear Effects}

After the preliminaries of the previous section we transfer our attention to a
single pendulum and moreover we use the second order approximation to estimate
the corrections related to a non-linear term in the motion equation. Thus if we
use same notations and techniques as before, we find the power spectrum density
$$S(f)={{4\xi\omega_0K_BT\over m}\over{[\omega_{0}^{2}(1+{3\alpha\over 2}-9
\alpha^2)-\omega^2]^2+(2\xi\omega_0\omega)^2[1-{27\over8}({\alpha\over\xi})^2]-
{(12\xi\alpha\omega_{0}^{3})^2\over{2(1-\xi^2)}}[\frac{\omega^2(\xi^2-3)+2
\omega_{0}^{2}\xi^2(1+8\xi^2)}{(\omega^2-\omega_{0}^{2}(1+8\xi^2))^2+
(6\xi\omega_0\omega)^2}+\frac{\xi^{-2}(2\omega^2-9\omega_{0}^{2}(1+\xi^2))}
{(\omega^2-9\omega_{0}^{2})^2+(6\xi\omega_0\omega)^2}]}}$$
with $\omega=2\pi f$ and $\alpha={2K_BT\epsilon\over(m\omega_{0}^{2})^2}$. From
the Boltzmann distribution in the non-linear case we easily recover the result
$$<q^n>=\frac{\int_{-\infty}^{\infty}\int_{-\infty}^{\infty}e^{-{1\over{K_BT}}
({p^2\over{2m}}+{1\over2}m\omega_{0}^{2}q^2+\epsilon{q^4\over4})}q^n\mbox{d}q
\mbox{d}p}{\int_{-\infty}^{\infty}\int_{-\infty}^{\infty}e^{-{1\over{K_BT}}
({p^2\over{2m}}+{1\over2}m\omega_{0}^{2}q^2+\epsilon{q^4\over4})}\mbox{d}q
\mbox{d}p}=(n-1)!!{K_BT\over{(m\omega_{0}^{2})^{n/2}}}[{1\over2}+\frac{e^{-2 n
\alpha}}{2(1+{3\alpha\over4})^{32\over3}}]^{n\over8}\quad.$$
This relation is a sort of scaling-law in the second order approximation. As an
example we may write
$$<q^4>-3<q^2><q^2>=-3\alpha({K_BT\over(m\omega_{0}^{2})})^2+14\alpha^2({3K_BT
\over{2m\omega_{0}^{2}}})^2$$
and compare it with
$$<q^4(t)>_{conn}=-3\alpha({K_BT\over{m\omega_{0}^{2}}})^2+14\alpha^2({3K_BT
\over{2m\omega_{0}^{2}}})^2\,.$$
This is the asymtpotic expansion of $<q^4(t)>_{conn}$ for $t\rightarrow\infty$.
In the same notation
$$<q^2>=({K_BT\over{m\omega_{0}^{2}}})\frac{\int_{-\infty}^{+\infty}e^{-({q'^2
\over2}+{\alpha\over 8}q'^4)}q'^2\mbox{d}q'}{\int_{-\infty}^{+\infty}e^{-({q'^2
\over2}+{\alpha\over 8}q'^4)}\mbox{d}q'}=({K_BT\over{m\omega_{0}^{2}}})(1-{3
\alpha\over2}+6\alpha^2)\quad.$$

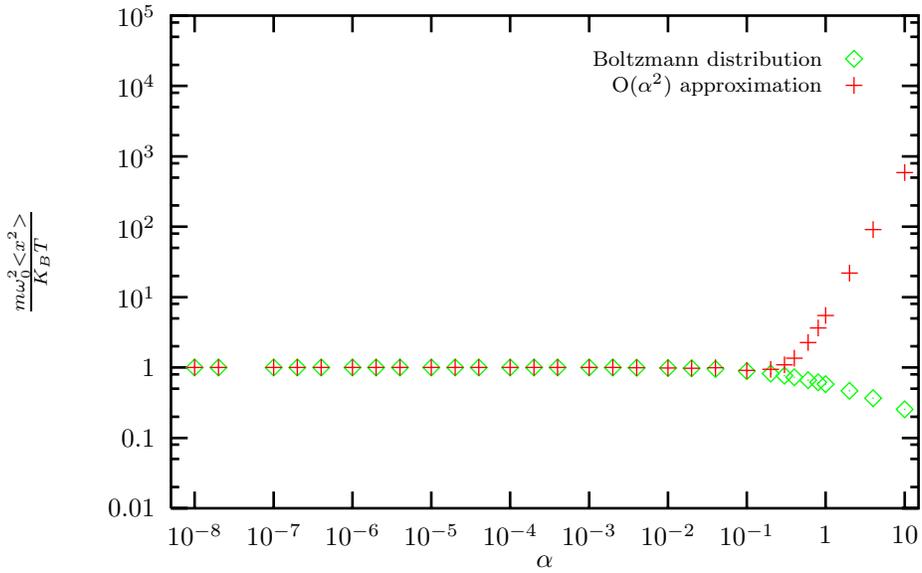
\begin{figure}
\begin{center}
% GNUPLOT: LaTeX picture with Postscript
\setlength{\unitlength}{0.1bp}
\begin{picture}(3600,2160)(0,0)
\put(3054,1846){\makebox(0,0)[r]{\footnotesize{$\mbox{O}(\alpha^2)$ approximation}}}
\put(3054,1946){\makebox(0,0)[r]{\footnotesize{Boltzmann distribution}}}
\put(2008,51){\makebox(0,0){$\alpha$}}
\put(100,1180){%
%
\makebox(0,0)[b]{\shortstack{$\frac{m\omega_{0}^{2}<x^2>}{K_BT}$}}%
%
}
\put(3365,151){\makebox(0,0){10}}
\put(3067,151){\makebox(0,0){1}}
\put(2770,151){\makebox(0,0){$10^{-1}$}}
\put(2473,151){\makebox(0,0){$10^{-2}$}}
\put(2176,151){\makebox(0,0){$10^{-3}$}}
\put(1878,151){\makebox(0,0){$10^{-4}$}}
\put(1581,151){\makebox(0,0){$10^{-5}$}}
\put(1284,151){\makebox(0,0){$10^{-6}$}}
\put(987,151){\makebox(0,0){$10^{-7}$}}
\put(689,151){\makebox(0,0){$10^{-8}$}}
\put(540,2109){\makebox(0,0)[r]{$10^5$}}
\put(540,1844){\makebox(0,0)[r]{$10^4$}}
\put(540,1578){\makebox(0,0)[r]{$10^3$}}
\put(540,1313){\makebox(0,0)[r]{$10^2$}}
\put(540,1047){\makebox(0,0)[r]{$10^1$}}
\put(540,782){\makebox(0,0)[r]{$1$}}
\put(540,516){\makebox(0,0)[r]{$0.1$}}
\put(540,251){\makebox(0,0)[r]{$0.01$}}
\end{picture}
\caption{The expansion in $\alpha$ leads to good approximations of $<x^2>$. The
approximated expression is not longer valid when $\alpha\rightarrow1$. There is
an asymptotic limit $<x^2>\sim\alpha^{-{1\over2}}K_BT(m\omega_{0}^{2})^{-1}$ if $\alpha$ grows}
\label{fig:cf}
\end{center}
\end{figure}

It is also possible to obtain a numerical evaluation of the involved integrals
and make a comparison of the result with the expansion in $\alpha$. One has for
$q^2$ two series of points as in Fig.\ref{fig:cf} for many choices of $\alpha$.

We put $q'=\sqrt{m\omega_{0}^{2}\over{K_BT}}q$ in the integrals. It's necessary
for the expansion to be valid that the value of $\alpha$ be small. For $\alpha$
great it's logical that the perturbative approximation is not longer justified.

We also know that
$$<q^2(t)>=\int_{-\infty}^{\infty}S(f)\mbox{d}f=({K_BT\over{m\omega_{0}^{2}}})
(1-{3\alpha\over2}+6\alpha^2)$$
from the corresponding dynamic correlation function recovered by (\ref{eq:sf}).

We are less interested in static properties than in dynamic properties, so turn
our attention to the power spectrum density. We may apply the previous analytic
estimations to a simple pendulum having length $l\sim1m$ and mass $m\sim500kg$.

Indeed even in a multi stage pendulum holding the mirrors of an interferometer,
the thermal noise depends on the last pendulum and on the test mass itself, and
above the pendulum proper frequency the dynamics mimics the response of a free
mass. In order to have an idea of the noise if some non-linearity arises we use
a value for $m$ staying between a single test mass and all the tilting filters
in the VIRGO superattenuator \cite{FD}.

Accordingly to the antennas' pendula requirements that are being built, we take
$l$ as the coarse order of magnitude for the length of the system's suspension.

This argument inspired the choice of the mechanical quality factor's value too.

In Fig.\ref{fig:sp} the PSD is shown. It gives the analysed system's behaviour.

Because of the non-linear contributions, there are two corrections particularly
peaked in correspondence of $2\pi f\sim{g\over l}$ and $2\pi f\sim3{g\over l}$.

Another consequence of non-linearity is related to the bins correlation and we
develop it in this section. Owing to white noise if the system were linear bins
would be completely uncorrelated.
On the contrary from the non-linear components of the response it follows that
$$<|\tilde q(f)|^2|\tilde q(f')|^2>_{conn}=-{6\epsilon\over{K_BT}}({\omega_0
\over\xi}-{2\pi^2(f^2+f'^2)\over{\xi\omega_0}})|S(f)|^2|S(f')|^2{1\over{2\eta}}
\quad.$$
Here $2\eta$ is the width of each bin. Thanks to the smallness of $K_BT$ it can
be found that the bins correlation is dramatically small. We report the related
plot in Fig.\ref{fig:bc} with the usual coefficients as before. If both $f$ and
$f'$ approach the resonance frequency the bins correlation shows its top value.

\section{Diagrams Representation}

As a companion to a perturbative calculation of the solution of the stochastic
differential equation, it is sometimes advantageous to consider the alternative
method of {\it stochastic diagrams}.

The solution of the linear differential equation is associated with a line and
a cross denoting the Green's function and the stochastic force respectively. If
the non-linear contribution is taken into account we may iteratively solve the
stochastic differential equation. The non-linear term contributes a vertex that
represents the convolution product of the imperturbed Green's function and the
solutions of the equation.

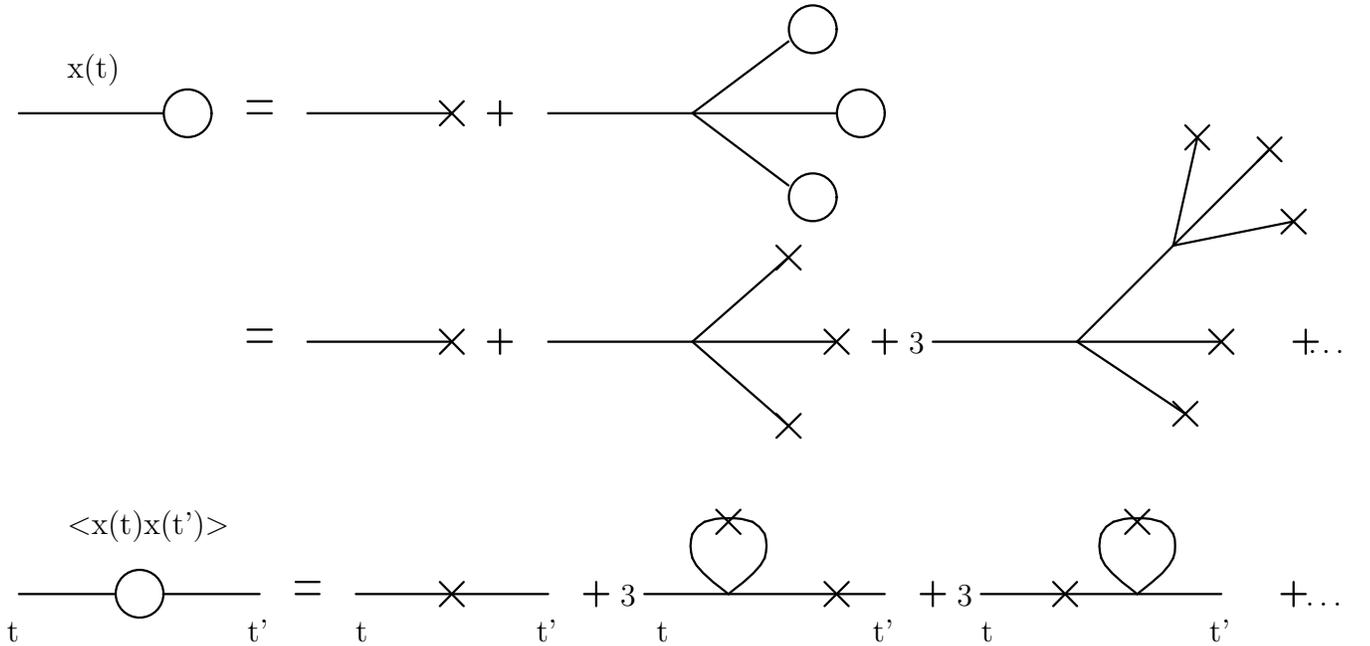
\begin{figure}[hb]
\begin{center}
\setlength{\unitlength}{0.00083333in}
\begingroup\makeatletter\ifx\SetFigFont\undefined
% extract first six characters in \fmtname
\def\x#1#2#3#4#5#6#7\relax{\def\x{#1#2#3#4#5#6}}%
\expandafter\x\fmtname xxxxxx\relax \def\y{splain}%
\ifx\x\y   % LaTeX or SliTeX?
\gdef\SetFigFont#1#2#3{%
  \ifnum #1<17\tiny\else \ifnum #1<20\small\else
  \ifnum #1<24\normalsize\else \ifnum #1<29\large\else
  \ifnum #1<34\Large\else \ifnum #1<41\LARGE\else
     \huge\fi\fi\fi\fi\fi\fi
  \csname #3\endcsname}%
\else
\gdef\SetFigFont#1#2#3{\begingroup
  \count@#1\relax \ifnum 25<\count@\count@25\fi
  \def\x{\endgroup\@setsize\SetFigFont{#2pt}}%
  \expandafter\x
    \csname \romannumeral\the\count@ pt\expandafter\endcsname
    \csname @\romannumeral\the\count@ pt\endcsname
  \csname #3\endcsname}%
\fi
\fi\endgroup
\begin{picture}(8187,4024)(0,-10)
\thicklines
\put(5025,3852){\ellipse{300}{300}}
\put(5325,3327){\ellipse{300}{300}}
\put(5025,2802){\ellipse{300}{300}}
\put(825,327){\ellipse{300}{300}}
\path(75,3327)(975,3327)
\path(1875,3327)(2775,3327)
\path(2700,3402)(2850,3252)
\path(2850,3402)(2700,3252)
\path(3375,3327)(4275,3327)
\path(4275,3327)(4875,3777)
\path(4275,3327)(5175,3327)
\path(4275,3327)(4875,2877)
\path(1500,3402)(1650,3402)
\path(1500,3327)(1650,3327)
\path(3000,3327)(3150,3327)
\path(3075,3402)(3075,3252)
\path(1875,1902)(2775,1902)
\path(2700,1977)(2850,1827)
\path(2850,1977)(2700,1827)
\path(3375,1902)(4275,1902)
\path(4275,1902)(5175,1902)
\path(4275,1902)(4875,2427)
\path(5100,1977)(5250,1827)
\path(5250,1977)(5100,1827)
\path(4800,2502)(4950,2352)
\path(4950,2502)(4800,2352)
\path(4275,1902)(4875,1377)
\path(4800,1452)(4950,1302)
\path(4950,1452)(4800,1302)
\path(5775,1902)(6675,1902)
\path(6675,1902)(7575,1902)
\path(6675,1902)(7275,2502)
\path(6675,1902)(7350,1452)
\path(7275,2502)(8025,2652)
\path(7275,2502)(7875,3102)
\path(7275,2502)(7425,3177)
\path(7500,1977)(7650,1827)
\path(7650,1977)(7500,1827)
\path(7275,1527)(7425,1377)
\path(7425,1527)(7275,1377)
\path(7950,2727)(8100,2577)
\path(8100,2727)(7950,2577)
\path(7350,3252)(7500,3102)
\path(7500,3252)(7350,3102)
\path(7800,3177)(7950,3027)
\put(1125,3327){\ellipse{300}{300}}
\path(7950,3177)(7800,3027)
\put(5925,252){\makebox(0,0)[lb]{\smash{{{\SetFigFont{12}{14.4}{rm}3}}}}}
\path(1500,1977)(1650,1977)
\path(1500,1902)(1650,1902)
\path(3000,1902)(3150,1902)
\path(3075,1977)(3075,1827)
\path(5400,1902)(5550,1902)
\path(5475,1977)(5475,1827)
\path(8025,1902)(8175,1902)
\path(8100,1977)(8100,1827)
\path(75,327)(675,327)
\path(975,327)(1575,327)
\path(2175,327)(3375,327)
\path(2700,402)(2850,252)
\path(2850,402)(2700,252)
\path(3975,327)(5475,327)
\path(5100,402)(5250,252)
\path(5250,402)(5100,252)
\path(4425,852)(4575,702)
\path(4575,852)(4425,702)
\path(6075,327)(7575,327)
\path(6525,402)(6675,252)
\path(6675,402)(6525,252)
\path(6975,852)(7125,702)
\path(7125,852)(6975,702)
\path(1800,402)(1950,402)
\path(1800,327)(1950,327)
\path(3600,327)(3750,327)
\path(3675,402)(3675,252)
\path(5700,327)(5850,327)
\path(5775,402)(5775,252)
\path(7950,327)(8100,327)
\path(8025,402)(8025,252)
\path(4500,327) (4449.964,365.607)
        (4407.507,399.901)
        (4342.691,458.186)
        (4300.280,507.128)
        (4275.000,552.000)

\path(4275,552) (4266.293,585.948)
        (4263.390,627.000)
        (4266.293,668.052)
        (4275.000,702.000)

\path(4275,702) (4306.695,745.305)
        (4350.000,777.000)

\path(4350,777) (4417.896,794.415)
        (4458.060,798.769)
        (4500.000,800.220)
        (4541.940,798.769)
        (4582.104,794.415)
        (4650.000,777.000)

\path(4650,777) (4693.305,745.305)
        (4725.000,702.000)

\path(4725,702) (4733.707,668.052)
        (4736.610,627.000)
        (4733.707,585.948)
        (4725.000,552.000)

\path(4725,552) (4699.720,507.128)
        (4657.309,458.186)
        (4592.493,399.901)
        (4550.036,365.607)
        (4500.000,327.000)

\path(7050,327) (6999.964,365.607)
        (6957.507,399.901)
        (6892.691,458.186)
        (6850.280,507.128)
        (6825.000,552.000)

\path(6825,552) (6816.293,585.948)
        (6813.390,627.000)
        (6816.293,668.052)
        (6825.000,702.000)

\path(6825,702) (6856.695,745.305)
        (6900.000,777.000)

\path(6900,777) (6967.896,794.415)
        (7008.060,798.769)
        (7050.000,800.220)
        (7091.940,798.769)
        (7132.104,794.415)
        (7200.000,777.000)

\path(7200,777) (7243.305,745.305)
        (7275.000,702.000)

\path(7275,702) (7283.707,668.052)
        (7286.610,627.000)
        (7283.707,585.948)
        (7275.000,552.000)

\path(7275,552) (7249.720,507.128)
        (7207.309,458.186)
        (7142.493,399.901)
        (7100.036,365.607)
        (7050.000,327.000)

\put(375,3552){\makebox(0,0)[lb]{\smash{{{\SetFigFont{12}{14.4}{rm}x(t)}}}}}
\put(0,27){\makebox(0,0)[lb]{\smash{{{\SetFigFont{12}{14.4}{rm}t}}}}}
\put(8100,252){\makebox(0,0)[lb]{\smash{{{\SetFigFont{12}{14.4}{rm}\ldots}}}}}
\put(8115,1827){\makebox(0,0)[lb]{\smash{{{\SetFigFont{12}{14.4}{rm}\ldots}}}}}
\put(1500,27){\makebox(0,0)[lb]{\smash{{{\SetFigFont{12}{14.4}{rm}t'}}}}}
\put(2175,27){\makebox(0,0)[lb]{\smash{{{\SetFigFont{12}{14.4}{rm}t}}}}}
\put(3300,27){\makebox(0,0)[lb]{\smash{{{\SetFigFont{12}{14.4}{rm}t'}}}}}
\put(4050,27){\makebox(0,0)[lb]{\smash{{{\SetFigFont{12}{14.4}{rm}t}}}}}
\put(5400,27){\makebox(0,0)[lb]{\smash{{{\SetFigFont{12}{14.4}{rm}t'}}}}}
\put(6075,27){\makebox(0,0)[lb]{\smash{{{\SetFigFont{12}{14.4}{rm}t}}}}}
\put(7500,27){\makebox(0,0)[lb]{\smash{{{\SetFigFont{12}{14.4}{rm}t'}}}}}
\put(375,702){\makebox(0,0)[lb]{\smash{{{\SetFigFont{12}{14.4}{rm}$<$x(t)x(t')$>$}}}}}
\put(5625,1827){\makebox(0,0)[lb]{\smash{{{\SetFigFont{12}{14.4}{rm}3}}}}}
\put(3825,252){\makebox(0,0)[lb]{\smash{{{\SetFigFont{12}{14.4}{rm}3}}}}}
\end{picture}
\caption{Graphical representation of $x(t)$ and $<x(t)x(t')>$. When the average
over the random force is taken, all crosses are joined together in all possible
ways. For example one can get $<x(t)x(t')>$ by combining two crosses at a point
in the corresponding product}
\label{fig:aa}
\end{center}
\end{figure}

This procedure is better illustrated by the first diagrams in Fig.\ref{fig:aa}.

The notation is meant to indicate the following. From the differential equation
we may write the solution as an iterative expansion. Accordingly $q(t)$ depends
on various products of noise terms which we average over to obtain correlation
functions. When there are several equivalent possible ways of combining crosses
a combinatorical factor arises.

The first order correction to the two-point correlation function is the result
of two tadpole-type diagrams. Their contribution is a correction to $\omega_0$.

In the second order approximation there are tadpole-type contributions as well.

Yet there are two-loops diagrams that introduce new terms with different poles
than the previous corrections.

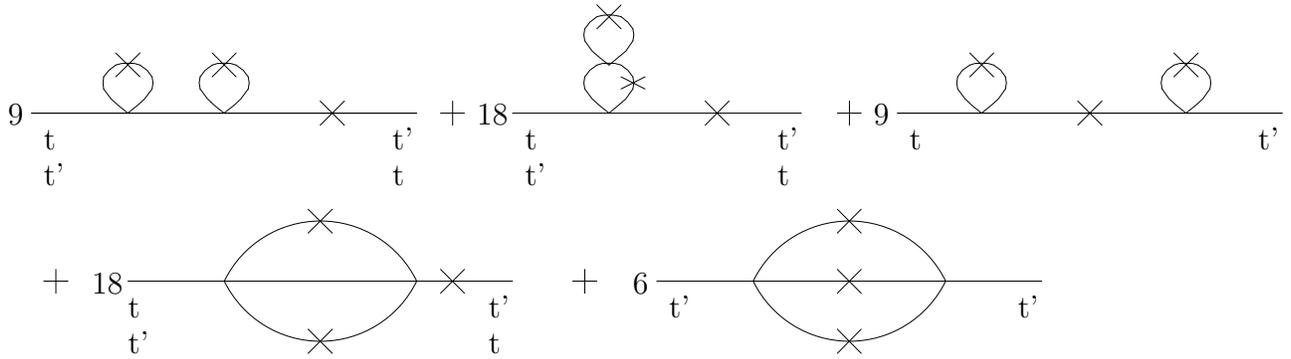
\begin{figure}[hb]
\begin{center}
\setlength{\unitlength}{0.00083333in}
\begingroup\makeatletter\ifx\SetFigFont\undefined%
\gdef\SetFigFont#1#2#3#4#5{%
  \reset@font\fontsize{#1}{#2pt}%
  \fontfamily{#3}\fontseries{#4}\fontshape{#5}%
  \selectfont}%
\fi\endgroup%
{\renewcommand{\dashlinestretch}{30}
\begin{picture}(7962,2229)(0,-10)
\put(5250.000,184.500){\arc{1335.000}{3.5952}{5.8296}}
\put(5250.000,769.500){\arc{1335.000}{0.4536}{2.6880}}
\path(4050,477)(6450,477)
\path(5175,177)(5325,27)
\path(5325,177)(5175,27)
\path(5175,927)(5325,777)
\path(5325,927)(5175,777)
\put(1950.000,184.500){\arc{1335.000}{3.5952}{5.8296}}
\put(1950.000,769.500){\arc{1335.000}{0.4536}{2.6880}}
\path(150,1527)(2550,1527)
\path(150,1527)(2550,1527)
\path(150,1527)(2550,1527)
\path(675,1902)(825,1752)
\path(825,1902)(675,1752)
\path(1275,1902)(1425,1752)
\path(1425,1902)(1275,1752)
\path(1950,1602)(2100,1452)
\path(2100,1602)(1950,1452)
\path(3150,1527)(4950,1527)
\path(3825,1752)(3975,1677)
\path(3975,1752)(3825,1677)
\path(3675,2202)(3825,2052)
\path(3825,2202)(3675,2052)
\path(4350,1602)(4500,1452)
\path(4500,1602)(4350,1452)
\path(5550,1527)(7950,1527)
\path(5550,1527)(7950,1527)
\path(6000,1902)(6150,1752)
\path(6150,1902)(6000,1752)
\path(7275,1902)(7425,1752)
\path(7425,1902)(7275,1752)
\path(6675,1602)(6825,1452)
\path(6825,1602)(6675,1452)
\path(5175,1527)(5325,1527)
\path(5250,1602)(5250,1452)
\path(2700,552)(2850,402)
\path(2850,552)(2700,402)
\path(750,477)(3150,477)
\path(2025,927)(1875,777)
\path(1875,927)(2025,777)
\path(2025,177)(1875,27)
\path(1875,177)(2025,27)
\path(225,477)(375,477)
\path(300,552)(300,402)
\path(3525,477)(3675,477)
\path(3600,552)(3600,402)
\path(5175,552)(5325,402)
\path(5325,552)(5175,402)
\path(2700,1527)(2850,1527)
\path(2775,1602)(2775,1452)
\path(750,1527) (688.338,1575.601)
        (645.128,1614.457)
        (600.000,1677.000)

\path(600,1677) (594.195,1714.500)
        (600.000,1752.000)

\path(600,1752) (631.695,1795.305)
        (675.000,1827.000)

\path(675,1827) (708.948,1835.707)
        (750.000,1838.610)
        (791.052,1835.707)
        (825.000,1827.000)

\path(825,1827) (868.305,1795.305)
        (900.000,1752.000)

\path(900,1752) (905.805,1714.500)
        (900.000,1677.000)

\path(900,1677) (854.872,1614.457)
        (811.662,1575.601)
        (750.000,1527.000)

\path(1350,1527)        (1288.338,1575.601)
        (1245.128,1614.457)
        (1200.000,1677.000)

\path(1200,1677)        (1194.195,1714.500)
        (1200.000,1752.000)

\path(1200,1752)        (1231.695,1795.305)
        (1275.000,1827.000)

\path(1275,1827)        (1308.948,1835.707)
        (1350.000,1838.610)
        (1391.052,1835.707)
        (1425.000,1827.000)

\path(1425,1827)        (1468.305,1795.305)
        (1500.000,1752.000)

\path(1500,1752)        (1505.805,1714.500)
        (1500.000,1677.000)

\path(1500,1677)        (1454.872,1614.457)
        (1411.662,1575.601)
        (1350.000,1527.000)

\path(3750,1527)        (3688.338,1575.601)
        (3645.128,1614.457)
        (3600.000,1677.000)

\path(3600,1677)        (3594.195,1714.500)
        (3600.000,1752.000)

\path(3600,1752)        (3631.695,1795.305)
        (3675.000,1827.000)

\path(3675,1827)        (3708.948,1835.707)
        (3750.000,1838.610)
        (3791.052,1835.707)
        (3825.000,1827.000)

\path(3825,1827)        (3868.305,1795.305)
        (3900.000,1752.000)

\path(3900,1752)        (3905.805,1714.500)
        (3900.000,1677.000)

\path(3900,1677)        (3854.872,1614.457)
        (3811.662,1575.601)
        (3750.000,1527.000)

\path(3750,1827)        (3688.338,1875.601)
        (3645.128,1914.457)
        (3600.000,1977.000)

\path(3600,1977)        (3594.195,2014.500)
        (3600.000,2052.000)

\path(3600,2052)        (3631.695,2095.305)
        (3675.000,2127.000)

\path(3675,2127)        (3708.948,2135.707)
        (3750.000,2138.610)
        (3791.052,2135.707)
        (3825.000,2127.000)

\path(3825,2127)        (3868.305,2095.305)
        (3900.000,2052.000)

\path(3900,2052)        (3905.805,2014.500)
        (3900.000,1977.000)

\path(3900,1977)        (3854.872,1914.457)
        (3811.662,1875.601)
        (3750.000,1827.000)

\path(6075,1527)        (6013.338,1575.601)
        (5970.128,1614.457)
        (5925.000,1677.000)

\path(5925,1677)        (5919.195,1714.500)
        (5925.000,1752.000)

\path(5925,1752)        (5956.695,1795.305)
        (6000.000,1827.000)

\path(6000,1827)        (6033.948,1835.707)
        (6075.000,1838.610)
        (6116.052,1835.707)
        (6150.000,1827.000)

\path(6150,1827)        (6193.305,1795.305)
        (6225.000,1752.000)

\path(6225,1752)        (6230.805,1714.500)
        (6225.000,1677.000)

\path(6225,1677)        (6179.872,1614.457)
        (6136.662,1575.601)
        (6075.000,1527.000)

\path(7350,1527)        (7288.338,1575.601)
        (7245.128,1614.457)
        (7200.000,1677.000)

\path(7200,1677)        (7194.195,1714.500)
        (7200.000,1752.000)

\path(7200,1752)        (7231.695,1795.305)
        (7275.000,1827.000)

\path(7275,1827)        (7308.948,1835.707)
        (7350.000,1838.610)
        (7391.052,1835.707)
        (7425.000,1827.000)

\path(7425,1827)        (7468.305,1795.305)
        (7500.000,1752.000)

\path(7500,1752)        (7505.805,1714.500)
        (7500.000,1677.000)

\path(7500,1677)        (7454.872,1614.457)
        (7411.662,1575.601)
        (7350.000,1527.000)

\put(2925,1452){\makebox(0,0)[lb]{\smash{{{\SetFigFont{12}{14.4}{rm}18}}}}}
\put(0,1452){\makebox(0,0)[lb]{\smash{{{\SetFigFont{12}{14.4}{rm}9}}}}}
\put(5400,1452){\makebox(0,0)[lb]{\smash{{{\SetFigFont{12}{14.4}{rm}9}}}}}
\put(525,402){\makebox(0,0)[lb]{\smash{{{\SetFigFont{12}{14.4}{rm}18}}}}}
\put(3900,402){\makebox(0,0)[lb]{\smash{{{\SetFigFont{12}{14.4}{rm}6}}}}}
\put(225,1302){\makebox(0,0)[lb]{\smash{{{\SetFigFont{12}{14.4}{rm}t}}}}}
\put(2400,1302){\makebox(0,0)[lb]{\smash{{{\SetFigFont{12}{14.4}{rm}t'}}}}}
\put(3225,1302){\makebox(0,0)[lb]{\smash{{{\SetFigFont{12}{14.4}{rm}t}}}}}
\put(4800,1302){\makebox(0,0)[lb]{\smash{{{\SetFigFont{12}{14.4}{rm}t'}}}}}
\put(5625,1302){\makebox(0,0)[lb]{\smash{{{\SetFigFont{12}{14.4}{rm}t}}}}}
\put(7800,1302){\makebox(0,0)[lb]{\smash{{{\SetFigFont{12}{14.4}{rm}t'}}}}}
\put(225,1077){\makebox(0,0)[lb]{\smash{{{\SetFigFont{12}{14.4}{rm}t'}}}}}
\put(2400,1077){\makebox(0,0)[lb]{\smash{{{\SetFigFont{12}{14.4}{rm}t}}}}}
\put(3225,1077){\makebox(0,0)[lb]{\smash{{{\SetFigFont{12}{14.4}{rm}t'}}}}}
\put(4800,1077){\makebox(0,0)[lb]{\smash{{{\SetFigFont{12}{14.4}{rm}t}}}}}
\put(750,252){\makebox(0,0)[lb]{\smash{{{\SetFigFont{12}{14.4}{rm}t}}}}}
\put(3000,252){\makebox(0,0)[lb]{\smash{{{\SetFigFont{12}{14.4}{rm}t'}}}}}
\put(6300,252){\makebox(0,0)[lb]{\smash{{{\SetFigFont{12}{14.4}{rm}t'}}}}}
\put(4125,252){\makebox(0,0)[lb]{\smash{{{\SetFigFont{12}{14.4}{rm}t'}}}}}
\put(750,27){\makebox(0,0)[lb]{\smash{{{\SetFigFont{12}{14.4}{rm}t'}}}}}
\put(3000,27){\makebox(0,0)[lb]{\smash{{{\SetFigFont{12}{14.4}{rm}t}}}}}
\end{picture}
}
\caption{Second order corrections for the two-point correlation function. There
are contributions including tadpoles. A tadpole is nothing but a contraction of
$x$ with itself at the same time. Since $<x(t)x(t)>$ is a constant tadpoles are
insertions of constant terms}
\label{fig:ab}
\end{center}
\end{figure}

Now observe the diagrams in Fig.\ref{fig:ab}. The first three ones contribute a
second order correction to $\omega_0$. The other two entail new terms which are
peaked near the resonance frequency and three times the resonance frequency. If
$\xi$ is large the former peak may not be distinguished from the principal one.

The latter is evident.

The elastic constant becomes
$$m\omega_{0}^{2}\rightarrow m\omega_{0}^{2}(1+{3\over2}\alpha-9\alpha^2)$$
and the damping constant
$$2m\xi\omega_0\rightarrow 2m\xi\omega_0(1-{27\alpha^2\over{16\xi^2}})\qquad.$$
In order to understand those modifications we may note that there are stronger
forces pulling the system back to the equilibrium configuration and the result
is that the measured recall constant $m\omega_{0}^{2}$ has a new greater value.

For linear systems in contact with a heat-bath it seems that the damping force 
have been expressly designed to constrain every motion below the thermal noise.

This will be true even in the non-linear case but due to the system's dynamics
and the appearance of up-converted proper frequencies damping times are longer.

This effect participates to the attenuation of the system's response below the
fundamental frequency as it is shown in Fig.\ref{fig:a2}. The coupling constant
assumes non-perturbative values and the numerical results confirm the tendency
of the perturbative case.

\section{Numerical Simulations}

In this section we are planning a numerical simulation to analyse the equation
\begin{equation}
\ddot q+2\xi\omega_0\dot q+\omega_{0}^{2}q=2\sqrt{K_BT\xi\omega_0\over m}F(t)
\end{equation}
with $D=2m\xi\omega_0K_BT$ and $<F(t)F(t')>=\delta(t-t')$ as previously stated.

For convenience we write
$$\left\{\begin{array}{l}\dot q={p\over m} \\ \dot p=-2\xi\omega_0p-m
\omega_{0}^{2}q+\sqrt{2D}F\end{array}\right.\quad\Phi=\left(\begin{array}{c} q
 \\ p \end{array}\right)$$
$$\dot\Phi_i=f_i(\Phi)+\Xi_i\quad<\Xi(t)_i\Xi(t')_j>=\delta(t-t')
\left(\begin{array}{cc} 0 & 0 \\ 0 & 4m\xi\omega_0K_BT\end{array}\right)$$
where $f_i(\Phi)$ is not derived from a potential. This may be immediately seen
from the cross derivative $\partial_if_j\neq\partial_jf_i$. The stochastic term
$\Xi$ is completely characterized by its zero mean and correlation matrix. This
latter one is singular.

If we look at the original equation of motion we see that its solution depends
both on the initial conditions and the distribution of the noise source $F(t)$.

As a consequence the statistical properties of the system are characterized by
$$P(q'',p'',t''|q',p',t')=<q'',p''|U(t'',t')|q',p'>$$
with $|q',p'>$ simultaneous eigenvector of $\hat q$ and $\hat p$. The evolution
operator is such that
$$U(t'',t')=e^{-\hat H(t''-t')}\quad\hat H=D\hat\Pi^{2}_{p}-i\hat\Pi_p(m
\omega_{0}^{2}\hat q+2\xi\omega_0\hat p)+{i\over m}\hat\Pi_q\hat p$$
and the following rules
$$[\hat\Pi_p,\hat p]=-i\quad[\hat\Pi_q,\hat q]=-i\qquad[\hat\Pi_p,\hat q]=[\hat
\Pi_q,\hat p]=[\hat\Pi_p,\hat\Pi_q]=[\hat p,\hat q]=0$$
are to be satisfied.

In other words $q$ and $p$ are regarded to as independent variables. At any $t$
their value is distributed according to the probability that a certain initial
configuration has evolved into another one because of a stochastic force whose
statistical properties are known.

We may also recover the distribution of every physical quantities depending on
$q$ and $p$. At every time the corresponding operator may be expressed by means
of the Heisenberg picture
$$\hat\Omega(\hat q(t'')\hat p(t''))=U^{-1}(t'',t')\hat\Omega(\hat q(t'),\hat 
p(t'))U(t'',t')\qquad{\mbox{d}\over{\mbox{d}t}}\hat\Omega(\hat q(t)\hat p(t))=
[\hat H,\hat\Omega(\hat q(t),\hat p(t))]$$
if the observable quantity $\Omega$ does not explicitly depend on $t$. Making a
comparison with the Poisson brackets of classical mechanics we note a reversed
sign. Utilizing the Baker-Campbell-Hausdorff formula in $U(t'',t')$ one obtains
\begin{eqnarray*}\lefteqn{
P(q'',p'',t''|q',p',t')=\int\mbox{d}q_1\int\mbox{d}q_2\int\mbox{d}p_1\int
\mbox{d}p_2<q'',p''|e^{-{t''-t'\over2}\hat H_1}|q_2,p_2>\times} \\ & & 
<q_2,p_2|e^{-(t''-t')\hat H_2}|q_1,p_2>
<q_1,p_2|e^{-{t''-t'\over2}\hat H_1}|q',p'>+\mbox{O}\left((t''-t')^3\right)
\end{eqnarray*}
with $\hat H=\hat H_1+\hat H_2$. If it happens that $m\omega_{0}^{2}=0$ we have
\begin{equation}
P(p'',t''|p',t')={1\over\sqrt{4\pi D\xi\omega_0(1-e^{-\xi\omega_0}(t''-t')
)}}\exp[-\frac{(p''-p'e^{-2\xi\omega_0(t''-t')})^2}{4D\xi\omega_0(1-e^{-\xi
\omega_0(t''-t')})}]\label{eq:pq}
\end{equation}
because of the identity
$$\exp[-(t''-t')(D\hat\Pi^{2}_{p}-i2\xi\omega_0\hat\Pi_p\hat p)]=\exp[-{D\over{
4\xi\omega_0}}\left(1-\exp[-\xi\omega_0(t''-t')]\right)\hat\Pi^{2}_{p}]
\times\exp[i2\xi\omega_0(t''-t')\hat\Pi_p\hat p]\qquad.$$
Next for the purpose of simplifying the algebra we write $t''-t'=\Delta t$. Now
we examine the situation specified by $m\omega_{0}^{2}\neq0$. Finally we derive
a three-steps process to $\mbox{O}(\Delta t)^2$ order. Successive higher orders
can also be obtained.

We first demonstrate that
\begin{eqnarray*}
<q'',p''|e^{-{\Delta t\over 2}\hat H_1}|q',p'>&=&{\delta(q''-q')\over\sqrt{2D
\pi\Delta t}}\exp\left(-\frac{(p''-p'+m\omega_{0}^{2}q'\Delta t)^2}{4D\Delta t}
\right) \\ \hat H_1&=&D\hat\Pi^{2}_{p}-im\omega_{0}^{2}\hat\Pi_p\hat q \\
<q'',p''|e^{-\Delta t\hat H_2}|q',p'>&=&\delta(q''-q'-{p'\over{2m\xi\omega_0}}
[1-\exp(-2\xi\omega_0\Delta t)])\, \delta(p''-p'\exp(-2\xi\omega_0\Delta t)) \\
\hat H_2&=&-i2\xi\omega_0\hat\Pi_p\hat p+{i\over m}\hat\Pi_q\hat p
\end{eqnarray*}
starting from showing that
\begin{eqnarray*}
<q'',p''|e^{-{\Delta t\over 2}\hat H_1}|q',p'>&=&<q'',p''|e^{-{\Delta t\over2}D
\hat\Pi^{2}_{p}}e^{i{\Delta t\over 2}m\omega_{0}^{2}\hat\Pi_p\hat q}|q',p'>=
<q'',p''|e^{-{\Delta t\over2}D\hat\Pi^{2}_{p}}|q',p'-m\omega_{0}^{2}q{\Delta
t\over 2}> \\
<q'',p''|e^{-\Delta t\hat H_2}|q',p'>&=&<q'',p''|q'(\Delta t),p'(\Delta t)> \\
\left\{\begin{array}{l}\left[\hat H_2,\hat q\right]={1\over m}\hat p\\ 
\left[\hat H_2,\hat p\right]=-2\xi\omega_0\hat p\end{array}\right.&&\Rightarrow
\qquad \left\{\begin{array}{ll}\dot q'(t)={1\over m}p'(t) & q'(0)=q' \\ 
\dot p'(t)=-2\xi\omega_0p'(t) & p'(0)=p'\end{array}\right.
\end{eqnarray*}
where it may be $m\omega_{0}^{2}q\rightarrow m\omega_{0}^{2}q+\epsilon q^3$. We
yield the following algorithm
$$\left\{\begin{array}{l}p_1=p'+{\Delta t\over2}f(q')+\sqrt{D\Delta t}\xi_1 \\
 q_1=q'\end{array}\right.\Rightarrow\left\{\begin{array}{l}p_2=(1-2
\xi\omega_0+{1\over 2}(2\xi\omega_0\Delta t)^2)p_1 \\ q_2=q_1+{p_1\over{2m\xi
\omega_0}}(2\xi\omega_0\Delta t-{1\over2}(2\xi\omega_0\Delta t)^2)
\end{array}\right.\Rightarrow\left\{\begin{array}{l}p''=p_2+{\Delta t
\over2}f(q_2)+\sqrt{D\Delta t}\xi_2 \\ q''=q_2\end{array}\right.$$
where $f(q)=m\omega_{0}^{2}q+\epsilon q^3$ but any $f(q)$ could be substituted.

With this method an algorithm of any precision may be planned. The fluctuations
of the physical quantities are very small due to $D=2m\xi\omega_0K_BT$. Because
of this we shall scale both the physical variables $p\rightarrow\sqrt{K_BTm}p$
and $q\rightarrow\sqrt{K_BT\over{m\omega_{0}^{2}}}q$ to get two variables with
mean square value one.

\section{Results and Conclusions}

In the following we shall use the same $\xi$ and $\omega_0$ previously adopted.

These are the only two important physical parameters because of the scaling we
introduced at the end of the previous section. The form we recovered for $S(f)$
is to be multiplied by the scaling factor ${K_BT\over{m\omega_{0}^{2}}}$. It is
convenient to recall that we wrote it as the spectrum at the zero order with a
term in the denominator standing for the ``self-energy''. It gives a correction
to the physical parameters.

Furthemore there is a brand new resonance appearing as an up-conversion of the
original single resonant frequency.

In Fig.\ref{fig:a1} the power spectrum density for $\alpha=0.1$ is shown. There
is a little peak at three times the resonance frequency as previously expected.

The results obtained from the numerical simulation are compared with the curve
referred to $\alpha=0.0$. It may be noted that $S(f)$ is reduced for low values
of $f$ and that the fundamental peak moves a little forward. Indeed we expected
this from the perturbative variations of $m\omega_{0}^{2}$ and $2m\xi\omega_0$.

The up-converted resonance is set in a frequency region corresponding to small
values of $S(f)$. This is due to the response of the system that behaves like a
filter attenuating those components whose frequency is greater than $\omega_0$.

For smaller values of $\alpha$ the up-conversion effect is not appreciable. Let
us discuss if $\alpha={2K_BT\epsilon\over(m\omega_{0}^{2})^2}<<1$. Actually the
quotient ${\epsilon\over(m\omega_{0}^{2})^2}$ is generally tremendously smaller
than the parameter $(2K_BT)^{-1}$. The conclusion is that the coupling constant
between non-linearities in the suspension system's behaviour and thermal noise
is so small that an enhancement of up-conversion effects is not realistic. This
is due to the smallness of thermal fluctuations and to the fact that there are
not contributions proportional to the coupling constant, because only tad-poles
diagrams contribute at the first order approximation. These diagrams contribute
a parameters' correction but do not modify the correlations at different times.

We expect that the resonant frequency should be greater if $\alpha$ goes up. It
happens if the system explores the non-linear zone of the phase space. When $T$
rises the greater amplitude of fluctuations just produce such an effect. If the
recall constant increases it is obvious that the opposite tendency is obtained.

Finally we note that formally there are two more peaks than in the linear case,
but one of these can not be distinguished from the fundamental one whose width
is proportional to $\xi$. This is related to nothing but the PSD's proper shape
and its typical parameters.

\appendix

\section{Evolution operator decomposition}

In this section we want to give a brief proof of the exactness of the identity
$$\exp[-(t''-t')(D\hat\Pi^{2}_{p}-i2\xi\omega_0\hat\Pi_p\hat p)]=\exp[-{D\over{
4\xi\omega_0}}\left(1-\exp[-\xi\omega_0(t''-t')]\right)\hat\Pi^{2}_{p}]
\times\exp[i2\xi\omega_0(t''-t')\hat\Pi_p\hat p]\qquad.$$
First of all we multiply the evolution operator by the identity such as to get
$$\exp[-(t''-t')(D\hat\Pi^{2}_{p}-i2\xi\omega_0\hat\Pi_p\hat p)]=\exp[-(t''-
t')(D\hat\Pi^{2}_{p}-i2\xi\omega_0\hat\Pi_p\hat p)]\,\exp[{D\over{4\xi\omega_0}
}\hat\Pi_{p}^{2}]\,\exp[-{D\over{4\xi\omega_0}}\hat\Pi_{p}^{2}]$$
and than we use
\begin{equation}
e^{-(t''-t')\hat H}\hat\Pi_{p}^{2}e^{(t''-t')\hat H}=\hat\Pi_{p}^{2}\exp\left(-
\xi\omega_0(t''-t')\right)\label{eq:h1}
\end{equation}
to write the form
\begin{equation}
e^{-(t''-t')\hat H}e^{{D\over{4\xi\omega_0}}\hat\Pi^{2}_{p}}e^{(t''-t')\hat H}
e^{-(t''-t')\hat H}e^{-{D\over{4\xi\omega_0}}\hat\Pi^{2}_{p}}=e^{{D\over{4\xi
\omega_0}}\hat\Pi^{2}_{p}\exp\left(-\xi\omega_0(t''-t')\right)}
e^{-(t''-t')\hat H}e^{-{D\over{4\xi\omega_0}}\hat\Pi^{2}_{p}} \label{eq:h2}
\end{equation}
with $\hat H=D\hat\Pi^{2}_{p}-i2\xi\omega_0\hat\Pi_p\hat p$. The commutation
relation $[\hat H,\hat\Pi_p]$ is sufficient to have the equation (\ref{eq:h1}).

It simply describes the temporal evolution in the Heisenberg picture. The final
identity is yielded using
$$\exp[{D\hat\Pi^{2}_{p}\over{4\xi\omega_0}}]\hat H\exp[-{D\hat\Pi^{2}_{p}\over{4\xi\omega_0}}]=\hat H-D\hat\Pi^{2}_{p}$$
inside the (\ref{eq:h2}). Substituting the expression for $\hat H$ we arrive at
the following simple identity
$$e^{{D\over{4\xi\omega_0}}\hat\Pi_{p}^{2}\exp\left(-\xi\omega_0(t''-t')\right
)}e^{-{D\over{4\xi\omega_0}}\hat\Pi_{p}^{2}}e^{{D\over{4\xi\omega_0}}
\hat\Pi^{2}_{p}}e^{-(t''-t')\hat H}e^{-{D\over{4\xi\omega_0}}\hat\Pi_{p}^{2}}=
e^{-{D\over{4\xi\omega_0}}\left(1-\exp[-\xi\omega_0(t''-t')]\right)
\hat\Pi^{2}_{p}}e^{i2\xi\omega_0(t''-t')\hat\Pi_p\hat p}$$
that is the outcome.

There are many other ways of demonstrating the above identity. Here we cite the
iterative way of developping the Baker-Campbell-Hausdorff formula in \cite{BC}.

\input{figure1}

\input{figure2}

%%%%%%%%%%%%%%%%%%%Figure of comparison%%%%%%%%%%%%%%%%%%%%%%%%%%%%%%%%%%%%%%%%
%%%%%%%%%%%%%%%%End of comparison%%%%%%%%%%%%%%%%%%%%%%%%%%%%%%%%%%%%%%%%%%%%%%

%%%%%%%%%%%%%%%%%%%%%%%%%%Beginning of expansion's figure%%%%%%%%%%%%%%%%%%%%%%
%%%%%%%%%%%%%%%%%%%%%%%%%%%End of expansion's figure%%%%%%%%%%%%%%%%%%%%%%%%%%%

%%%%%%%%%%%%%%%%%%%%%%Beginning of diagrams%%%%%%%%%%%%%%%%%%%%%%%%%%%%%%%%%%%
%%%%%%%%%%%%%%%%%%%%%%%%%%%%%%%%End of diagrams%%%%%%%%%%%%%%%%%%%%%%%%%%%%%%%%

\input{figure5a}

\input{figure5b}
\end{document}